\documentstyle[preprint,revtex]{aps}
\def\beq{\begin{eqnarray}}
\def\eeq{\end{eqnarray}}
\begin{document}

\begin{title}
\begin{center}
 The Haldane Energy Gap\\
of A Doped Linear-Chain Heisenberg Antiferromagnet
\end{center}
\end{title}

\author{ Zhong-Yi Lu,$^a~~~~~~$   Zhao-Bin Su,$^a~~~~~~$    Lu Yu$^{a,b}$}
\begin{instit}
$^a$Institute of Theoretical Physics, Academia Sinica, \\
P. O. Box 2735, Beijing 100080, China \\
$~~~~$\\
$^b$International Centre for Theoretical Physics, 34100 Trieste, Italy \\
\end{instit}

\begin{abstract}

Using the valence-bond-solid (VBS) approach and the Schwinger boson mean
field approximation, we study the dependence of the Haldane gap
of a spin-1 linear chain Heisenberg antiferromagnet on impurity doping  with
different spins.  The impurity spins affect the singlet pairing
order parameter $\Delta $ and the
constraint factor $\lambda$. As a result, the Haldane gap is reduced by
a factor $ \sim n_i^{2/3}$,
with $n_i$ as the impurity concentration, and
  eventually collapses at $n_i \sim 1/\xi$ with  $\xi$ as the VBS
correlation length. This theoretical prediction can be verified
by neutron scattering experiments.
\end{abstract}

PACS  numbers:  75.30.Ds,75.30.Hx,75.50.Ee

\newpage

Some years ago Haldane \cite{Haldane} conjectured that the excitation
spectrum of
a linear-chain Heisenberg antiferromagnet (LCHA) with integer spin
has a finite energy gap $E_H$ above its singlet ground state,
while a LCHA with half-integer spin has a gapless spectrum.
This conjecture has been strongly supported by an exact solution of a
specific model\cite{aklt}, numerical studies\cite{Botet},\cite{Liang}
and experiments\cite{Buyers},\cite{ren}.  Recently, there has been a revived
interest in this problem due to the experimental observation of spin 1/2
degrees  of freedom\cite{hagi},\cite{gla} at the ends of a finite spin-1
chain induced by doping with different spins,
as predicted by  theory\cite{aklt},\cite{ken}.

P.W. Anderson\cite{and} has proposed the Resonant-Valence-Bond (RVB) model
to describe some spin 1/2 antiferromagnet systems. Affleck, Kennedy, Lieb
and Tasaki (AKLT) \cite{aklt} have generalized this idea to spin-1 LCHA
by proposing the Valence-Bond-Solid (VBS) state. In this VBS state, the
valence bonds formed by two 1/2 spins as a singlet
$\uparrow\downarrow-\downarrow\uparrow$,
 connect two nearest neighbors, while the two 1/2 spins on the same site
should be symmetrized to form a triplet state $S=1$.
 This is a translationally
invariant, singlet state. The spin correlation functions decay exponentially
and there is a gap $E_H$ in the energy spectrum which  is closely
related to the VBS order parameter.

Consider a generalized Heisenberg Hamiltonian
\beq\label{H0}
H_{0}=J_0\sum_i\left[ \vec{S}_i\cdot\vec{S}_{i+1}-\beta(\vec{S}_i\cdot
\vec{S}_{i+1})^2\right],
\eeq
\noindent where $-1\leq \beta \leq 1$ is a parameter.  The VBS state proposed
by AKLT is an exact, non-degenerate  ground state of (\ref{H0}) for $\beta=
-1/3$.  The theoretical analysis using the non-Abelian bosonization technique
and the conformal field theory\cite{aff}, as well as the exact diagonalization
for small clusters\cite{ken} seems to show that systems described by (\ref{H0})
 with $-1<\beta <1$ belong to the same universality class as the $\beta=
-1/3$ case, so its properties are generic to all systems of this
class. Recently, Ng\cite{ng}
has applied  the Schwinger boson mean field approximation (SBMFA)
adopted by Arovas and Auerbach\cite{aa}  for quantum spin systems
to study the end states in a spin-1 chain.
His results
agree semi-quantitatively with those of exact diagonalization.  The advantages
and weakness of the SBMFA have been discussed in his paper and will
not be repeated here.

In this Letter we use the SBMFA
to consider the effect of random doping with different spins on the Haldane
gap.
In this approach the Haldane
gap $E_H$  is determined by the VBS order parameter $\Delta$
and the single-occupancy
constraint factor $\lambda$.  Upon doping, apart from the end states,  $\Delta$
and $\lambda$ are modified
by both spin-flip scattering and the difference in the spin values.  By
solving self-consistently the system of equations for the order parameter
and the constraint we find that the reduction of the Haldane gap $E_H$
originating from the VBS structure is proportional to $n_i^{2/3}$, where
$n_i$ is the impurity concentration, and it collapses when $n_i$ is comparable
to $\xi^{-1}$, the inverse of the VBS correlation length.  The
situation here is rather similar to superconductors
doped with paramagnetic impurities which break the time-reversal symmetry
of the singlet pairing, giving rise to reduction and eventual collapsing
of the energy gap\cite{ag}.  Of course, if the spin chain is strictly
one-dimensional, doping by non-magmetic impurities will break the chain.
However, the real system is only quasi-one-dimensional, so the interchain
coupling as well as the superexchan
The dynamic structure factor $S(q, \omega)$, measured in the neutron scattering
experiments, is the Fourier transform of the spin-spin correlation function
$<\vec{S}_i(t)\cdot \vec{S}_j(0)>$, where $< \cdot \cdot \cdot>$ means thermal
average.  Apart f

The Hamiltonian of a LCHA doped with impurity spins can be written
as\cite{wolf}:
\begin{equation}\label{51}\begin{array}{ccc}
H=H_{0}+H^{im},&
H_{0}=J_0\sum_i\vec{S}_i\cdot\vec{S}_{i+1},&H^{im}=\sum_{<i\alpha>}\vec{A}_
{\alpha}\cdot\vec{S}_i,\\[3mm]
\vec{A}_{\alpha}=g\vec{S}_{\alpha}^{im}-J_0\vec{S}_{\alpha},&
\vec{S}^2=S(S+1),&J_0>0,\end{array}\end{equation}
 where $\vec{S}^{im}$ is the impurity spin, $g$ is the coupling
between
impurity and host spins, $\alpha$ denotes impurity site, $<i\alpha>$ means
summation over all impurity spins with
$i=\alpha\pm 1$.

We will discuss a generic case when
impurities are randomly distributed
along the LCHA and the averages are
$\overline{\vec{S}^{im}}=0$, $\overline{\vec{S}_{\theta}^{2~im}}=
1/3~S^{im}(S^{im}+1)$(where $\theta$ means any direction),
 $\bar{\vec{A}^2}=g^2S^{im}(S^{im}+1)+J_0^2S(S+1).$

The spin $\vec{S}$ can be represented
by Schwinger-Boson-operators as \cite{aa},
$\hat{S}^+=a^{\dagger}b,~\hat{S}^-=ab^{\dagger},~
\hat{S}^z=1/2(a^{\dagger}a-b^{\dagger}b),~
a^{\dagger}a+b^{\dagger}b=2S$, where $a,a^{\dagger},b,b^{\dagger}$ are boson
operators, $[a,b]=
[a,b^{\dagger}]=0,~ [a,a^{\dagger}]=1.$
It is convenient to
use
Nambu's four-component formalism to treat the Hamiltonian (\ref{51}).
Let $\psi_i=(a_i,~ b_i,~b_i^{\dagger},~a_i^{\dagger})^T$,
$\psi(k)=(a_k,~b_{k},~b_{-k}^{\dagger},~a_{-k}^{\dagger})^T$,
then $\psi_i^{\dagger}=(a_i^{\dagger},~b_i^{\dagger},~b_i,~a_i)$,
$\psi^{\dagger}(k)=(a_k^{\dagger},~b_{k}^{\dagger},~b_{-k},~a_{-k})$, where
$T$ means transposition and $k$ is the wave vector.
To carry out the decoupling
in the SBMFA we introduce the VBS  order parameter
$\Delta_{ij}=<a_ib_j-b_ia_j>$, with
$ij$ denoting the nearest neighbor pair of sites.

We first  consider the  undoped system.  In the SBMFA the Hamiltonian can be
written as
\begin{eqnarray} \label{haml}
&&H_0  = \sum_{i}~ \psi_{i}^{\dagger}~M_{i,i+1}~\psi_{i,i+1} +
\sum_{i}~\psi_{i}^{\dagger}~\Lambda_{i}~\psi_{i}  \nonumber \\ &&~~~~~+\sum_{i}
J\vert \Delta_{0~i,i+1}\vert^2+\sum_i\lambda_{0i}(-1-2S)~,
\end{eqnarray}
\noindent
where
$$
M_{i,i+1}  =  J
\left( \begin{array}{cc}0&-\Delta_{0~i,i+1} \sigma_3
\\   \Delta_{0~i,i+1}^{*} \sigma_3&0 \end{array} \right) ,~~~
 \Lambda_i  = \frac{\lambda_{0i}}{2}
\left( \begin{array}{cc}I & 0\\
0&I \end{array}\right) ,
$$
while $*$ means complex-conjugate,  $J=J_0/2$, $I$ and $\sigma_3$ are
$2\times 2$ unit and  Pauli matrix, respectively.

Performing the gauge transformation of Read and Newns
\cite{Read}  to absorb the phase factor and to remove the redundancy of
constraint, the Lagrangian corresponding to eq.(\ref{haml}) can be written in
the momentum space as
\begin{equation}\label{lag}
L=1/2\sum_k\psi^{\dagger}(k)(\omega\Omega_1-\lambda_0\Omega_2
+J\Delta_0\nu_k\Omega_3)\psi(k)+const,
\end{equation}
where $\nu_k=2\sin k$, $\Delta_0$ is the module of the VBS order
parameter.  Here we take the
lattice spacing $a$ as the length unit  and
denote $\vec{\Omega}=(\Omega_1,\Omega_2,\Omega_3)$ as
$$\Omega_1=\left(\begin{array}{cc} I&0\\0&-I\end{array}\right),\Omega_2=
\left(\begin{array}{cc}I&0\\0&I\end{array}\right),\Omega_3=
\left(\begin{array}{cc}0&\sigma_3\\ \sigma_3&0\end{array}\right).$$

{}From eq.(\ref{lag}) one obtains the boson Green function
\begin{equation}\label{green}
\displaystyle G_0(k,\omega)=
\displaystyle \frac{\omega\Omega_1+\lambda_0\Omega_2+
J\Delta_0\nu_k\Omega_3}{\omega^2-(\lambda_0 ^2-4J^2\Delta_0^2\sin^2k)+i\eta},
\end{equation}
where $\eta$ is a positive infinitesimal.
The Schwinger boson excitations
 form a continuum band and their dispersion
relation is $\epsilon(k)=(\lambda_0^2-4J^2\Delta_0^2\sin^2k)^{1/2}$.

Now we can easily calculate the dynamic structure factor
at  $T=0K$ as follows
\begin{equation}\begin{array}{l}\label{spin} \displaystyle
S(q,\omega)=12\pi\int~\frac{dk_1}{2\pi}~\int~\frac{d\omega_1}{2\pi}~
\int~\frac{d\omega_2}{2\pi}~\delta(\omega-\omega_1-\omega_2)\\[2mm]
\displaystyle
\{{\rm Im}G_{044}(k_1,\omega_1){\rm Im}G_{011}(q+k_1,\omega_2)
 -{\rm Im}G_{042}
(-k_1,\omega_1){\rm Im}G_{013}(q+k_1,\omega_2)\},
\end{array}\end{equation}
where $G_{044}$,$G_{011}$,$G_{042}$ and $G_{013}$ are  elements of the Green
function matrix (\ref{green}).
Integrating over  $\omega_1$ and $\omega_2$ of eq.(\ref{spin}), we find that
the $\delta$-function representing
the energy conservation becomes
$\delta(\omega-(\lambda_0^2-4J^2
\Delta_0^2\sin^2 k_1)^{1/2}-(\lambda_0^2-4J^2\Delta_0^2\sin^2(q+k_1))^{1/2}).$
Hence only when $\omega\geq 2(\lambda_0^2-4J^2
\Delta_0^2)^{1/2}$, a neutron can be scattered and
there must be two  Schwinger bosons to form  a physical magnetic
excitation for spin $S=1$. It is also clear from the Schwinger-Boson
constraint for the spin-1 chain
($a^{\dagger}a+b^{\dagger}b=2$) that
they must appear in pairs. This is entirely consistent
with the description based on field-theoretical mappings\cite{Affleck}.
Thus the Haldane gap $E_{H_0}$ is the bottom of
the spin excitation continuum and
is given by  $E_{H_0}=2(\lambda_0^2-4J^2\Delta_0^2)^{1/2}$.
The parameters $\Delta_0$
and  $\lambda_0$ can be
consistently evaluated  \cite{ng},\cite{lu}.
At temperature $T=0K$, they are $\lambda_0=4.7296J,~\Delta_0=2.36312,~
E_{H_0}=0.1772J_0.$

Now we consider the doped system.
In the following $\vec{A}_{\alpha}$
is referred to as a quasiclassical quantity.
In the Nambu representation
the impurity part of the Hamiltonian (\ref{51}) can be written as
$\displaystyle
H^{im}=1/4 \sum_{<i\alpha>}\psi_i^{\dagger}
(\vec{A}_{\alpha}\cdot\vec{\Sigma})\psi_i, $
in which $\vec{\Sigma}=(\Sigma^1,\Sigma^2,\Sigma^3)$
and
$$\begin{array}{ccc}
\Sigma^1=\left(\begin{array}{cc} \sigma_1&0\\[2mm]
0&\sigma_1\end{array}\right),&\Sigma^2=\left(\begin{array}{cc}
\sigma_2&0\\[2mm] 0&\sigma_2 \end{array}\right),
&\Sigma^3=\left(\begin{array}{cc}
\sigma_3&0\\[2mm] 0&-\sigma_3 \end{array}\right),
\end{array}$$
with $\sigma_1$, $\sigma_2$, $\sigma_3$ as Pauli matrices.

Since the impurities are randomly distributed, their average effect is zero if
each impurity scatters a Schwinger boson only
once. Thus the multiple-scattering
effect should be considered. Analysing the Feynman graphs, one finds
that the contributions from the crossing diagrams  are very small and may be
neglected.   The diagrams are similar to those of the impurity scattering in
the
 superconductors \cite{ag},
 and we will not describe the details here. The Green function
$\bar{G}(k,\omega)$ of the doped system
 after averaging over the random distribution recovers the translational
invariance.
Hence we have
$\bar{G}^{-1}(k,\omega)=\bar{G}_0^{-1}(k,\omega)-\Sigma(k,\omega)$,
where $\Sigma(k,\omega)$ is the self-energy term due to the impurity
scattering and $\bar{G}_0^{-1}(k,\omega)=\omega\Omega_1-\lambda\Omega_2
+J\Delta\nu_k\Omega_3$,
which has the same form as eq.(\ref{green}).  However, parameters
$\lambda$ and $\Delta$ should be self-consistently determined like
$\lambda_0$ and $\Delta_0$ and
in principle they are
different from the latter due to the impurity effect.
Furthermore, we make an ansatz that the inverse of
the Green function $\bar{G}^{-1}(k,z)=\bar{Z}z\Omega_1
-\bar{\lambda}\Omega_2+J\bar{\Delta}
\bar{Z}\nu_k\Omega_3$,
where $z$ is the frequency  in the complex plane,
$\bar{Z}$ is the wave function renormalization factor,
$\bar{\lambda}$ and $\bar{\Delta}$ are  the renormalized parameters for
the $\lambda$-multiplier and the order parameter, respectively.
All of them are functions of $k$ and $z$.
The
self-energy part after averaging over random distribution can be written
as
\begin{equation}\label{l31}\displaystyle
\Sigma(k,\omega)=n_i\int (\vec{A}\cdot\vec{\Sigma})^2
\bar{G}(k_1,\omega)
\cos^2(k-k_1)\frac{dk_1}{2\pi}.
\end{equation}
{}From the definitions given above and eq.(\ref{l31}), we obtain  a closed set
of integral equations
to determine $\bar{Z},~\bar{\lambda},$ and  $ \bar{\Delta}.$
Since the doping density $n_i$ is rather small, they can be expanded in powers
of $n_i$ and
we will keep only  the leading order of $n_i$ in the following.
Accordingly we obtain that \cite{lu}
 \begin{equation}\label{ren}
\bar{Z}z=z(1-\tau(k,z)),~~~\bar{\lambda}=\lambda(1+\tau(k,z)),~~~
\bar{\Delta}\bar{Z}=\Delta(1+\tau(k,z)) ,
\end{equation}
where
$$\tau(k,z)=\displaystyle \frac{n_i\bar{\vec{A}^2}}{4}\left\{
\displaystyle (4\cos^2k+\frac{(z^2-\lambda^2)\cos2k}{J^2\Delta^2})
(z^2-\lambda^2)^{-1/2}(z^2-\lambda^2
+4J^2\Delta^2)^{-1/2}-\frac{\cos 2k}{J^2\Delta^2}\right\}.$$
In this case  $\lambda$ and
$\Delta$ should  also  be  evaluated consistently.
If the impurity  density $n_i$ is zero, one  recovers $\lambda_0$
and $\Delta_0$. When $n_i$ is very small,
they can be expanded  up to $n_i$.

Now  we consider the  density  of the Schwinger boson
states for the doped system  given by
\begin{equation}\label{den} \displaystyle{
\frac{dN}{dz}=-\frac{1}{4 \pi}\int {\rm Tr}~
{\rm Im}~\bar{G}(k,z)\frac{dk}{2\pi}},
\end{equation}
where $N$ is the total number of sites and  Tr is the trace operation
in the Nambu representation.
We substitute eq.(\ref{ren}) into eq.(\ref{den}) and find  a critical value
\begin{equation}\label{mini}\displaystyle
\omega_{min}\displaystyle
=\frac{1}{2}E_{H}\left(1- \displaystyle
\frac{2}{E_H}~\left[\frac{(\bar{\vec{A}^2})^2E_H}{4J^2\Delta^2}
\right]^{1/3}
n_i^{2/3}+\frac{\bar{\vec{A}^2}}{3J^2\Delta^2}n_i\right),
\end{equation}
and
${dN}/{dz}$ is zero for $\displaystyle \omega<\omega_{min}$. Here
$E_H=2(\lambda^2-4J^2\Delta^2)^{1/2}.$

In SBMFA, the dynamic structure factor $S(q,\omega)$ is a two-particle
Schwinger-
boson Green function. The random doping effect consists of two parts.
The first is included in the simple product of two  single-particle
Schwinger-boson
Green functions $\bar{G}(k,\omega)$. The second is
the correlation effect between two Schwinger bosons induced by scattering at
the same impurity. This correlation effect
is mostly contributed by the two-particle Green function's
ladder diagrams and maximum crossing diagrams. Their contribution is at least
proportional to the impurity density $n_i$.
In the weak doping case we may ignore this effect.
Within this approximation $\bar{S} (q,\omega)$ in the presence of impurity
scattering can be easily calculated by replacing $G$ in eq. (\ref{spin}) with
$\bar{G}$ renormalized by impurity scattering.  As in the undoped case,
$S(q,\omega)$ vanishes for
is  the renormalized Haldane  gap
of the doped sample.
If we keep eq.(\ref{mini}) up to the first order of $n_i$,
$E_H$ and $\Delta$
can  be replaced by $E_{H_0}$ and $\Delta_0$, respectively.
Now we make a numerical estimate for the case of non-magnetic doping.
Such an experiment has been performed by Glarum $et~al.$ \cite{gla}
For  such  non-magnetic case we have $\bar{\vec{A}^2}=
J_0^2S(S+1)=8J^2$ (for $S=1$ ).
Substituting the parameter values for $\lambda_0$ and $\Delta_0$  into
(\ref{mini}), we find that
\begin{equation}\label{numer}
\bar{E}_H=E_{H_0}(1-5.6722n_i^{2/3}+0.4775n_i)
\end{equation}
from which we see that $\bar{E}_H\approx 0$ when $n_i\approx 7.8\%$.
In the SBMFA treatment
the VBS state's correlation length
$\xi$ is about 10 lattice spacings\cite{ng}, \cite{aa},
i.e., almost twice the exact value \cite{Liang}.
Within the same approximation we find that the Haldane
gap collapses when the correlation length is comparable with the average
impurity distance. This is reasonable and could be
expected from our argument at the very beginning.

The doping dependence of the Haldane gap
can be observed directly in neutron scattering experiments,
for example, in NENP doped with Cu.
It  will also show up in other physical quantities, e.g., the susceptibility.
We have evaluated these quantities in the doped case and  they will be
published later \cite{lu}.To our knowledge,
no data on this explicit dependence are available up to now. However, it
has been observed that the spin 1/2 ESR signal decays very rapidly at
much lower temperatures than the Haldane gap itself\cite{hagi},\cite{gla}.
  Mitra, Halperin and Affleck\cite{mha} have suggested to explain this
effect by considering the thermal excitations which will shift the ESR
frequency out of the observation window for short and intermediate chains,
 while broadening the signal for very long chains. This effect should
be there. However, we would suggest that there is a significant reduction
of the Haldane gap itself  (about 30 \%)
even at  1 \% doping by non-magnetic
impurities due to the effect considered in this Letter.  It seems to us
that these two complementary aspects should be  both taken into account
while interpreting the data.   Our calculation was done for zero temperature,
but it can be easily extended to finite temperatures to compare with actual
experiments\cite{lu}.  It is known \cite{temp} that the Haldane gap for undoped
systems first incr

In conclusion  we have considered the effects of doping by impurity spins
in LCHA on the Haldane gap.  In this calculation we have considered only
the impurity effect on the Schwinger boson
continuum spectrum which in turn leads
to reduction of the Haldane gap itself, like the early calculation
of Gor'kov and Abrikosov\cite{ag}
for the paramagnetic impurities in superconductors.
  However, there are bound states in both spin chain problem \cite{ken},
\cite{ng}, and superconductors\cite{yu}. The bound states in
superconductors are rather shallow
and it is difficult to observe them experimentally.
  On the contrary, the bound states in the LCHA are rather deep, very
close to the ground state.  Hybridizing with the ground state, they give
rise to the spin 1/2 signal observed in experiments.  They are
localized within the correlation length and in the low low doping case
 we can still focus on the scattering states. On the other hand,
if we discuss the collapse of the Haldane gap upon doping
quantitatively, both tail states of the continuum band and the band formed
by these localized states should be  taken into account.
In this case the impurity density $n_i$ is not very small any longer and we
cannot neglect the correlation effect between the Schwinger
bosons coming from scattering on the same impurity.  In fact, the
properties of such gapless LCHA with nonvanishing  VBS order parameter
should be very interesting to study.  Furthermore, we have treated the
impurity spins in this paper quasi-classically, but they are of quantum
nature, like the Kondo impurity in the normal metal and superconductors.
Their quantum fluctuations should have important effects.  The work on
these issues is in progress and will be reported later.

\vspace{3ex}
\centerline{\large\sf  Acknowledgements}\vspace{1ex}
We would like to thank  P. Mitra, T.K. Ng  and G.-M. Zhang for helpful
discussions.
 One of the authors (ZYL) is indebted to Drs. J.-X.Wang , J.-G. Hu and
S.-J. Qin  for their help. This work was supported in part
by the National Natural Science Foundation of China and China Center
of Advanced Science and Technology.

\end{document}